\documentclass{article}
\usepackage{spconf}
\usepackage{amsmath}
\usepackage{graphicx}

\usepackage{booktabs}
\usepackage{multirow}
\usepackage{rotating}

\usepackage[unicode=true,pdfusetitle,%
 pdfauthor={Siyang Wang, Gustav Eje Henter, Joakim Gustafson, Éva Székely},%
 pdfkeywords={speech synthesis, self-supervised speech representation, spontaneous speech},%
 bookmarks=true,bookmarksnumbered=true,bookmarksopen=false,%
 breaklinks=true,pdfborder={0 0 0},backref=false,colorlinks=true,citecolor=blue]{hyperref}

\usepackage[shortlabels]{enumitem} % Better control of lists and their spacing
\setlist{nolistsep} % Removes vertical spacing in lists
\title{A Comparative Study of Self-Supervised Speech Representations\\{}in Read and Spontaneous TTS}
\name{Siyang Wang, Gustav Eje Henter, Joakim Gustafson, Éva Székely\thanks{This research was supported by the Digital Futures project Advanced Adaptive Intelligent Systems (AAIS), the Swedish Research Council projects Connected (VR-2019-05003), Perception of speaker stance (VR-2020-02396), the Riksbankens Jubileumsfond project CAPTivating (P20-0298), and by the Wallenberg AI, Autonomous Systems and Software Program (WASP) funded by the Knut and Alice Wallenberg Foundation.}}

\address{Division of Speech, Music and Hearing, KTH Royal Institute of Technology, Stockholm, Sweden}
\begin{document}
%\ninept
%
\maketitle
\begin{abstract}
% Text-to-speech is an underexplored downstream task for self-surpversied learning (SSL) speech representations. ??
%Text-to-speech represents a promising downstream task for self-supervised learning (SSL) speech representations, which has not been explored sufficiently thus far.
%The potential of text-to-speech as a downstream task for self-supervised learning (SSL) speech representations deserves more attention and investigation.
% Current state-of-the-art neural TTS is dominated by a two-stage pipeline with mel-spectrogram as representation medium between the two stages. 
Recent work has explored using self-supervised learning (SSL) speech representations such as wav2vec2.0 as the representation medium in standard two-stage TTS, in place of conventionally used mel-spectrograms. It is however unclear which speech SSL is the better fit for TTS, and whether or not the performance differs between read and spontaneous TTS, the later of which is arguably more challenging.
This study aims at addressing these questions by testing several speech SSLs, including different layers of the same SSL, in two-stage TTS on both read and spontaneous corpora, while maintaining constant TTS model architecture and training settings. 
Results from listening tests show that the 9th layer of 12-layer wav2vec2.0 (ASR finetuned) outperforms other tested SSLs and mel-spectrogram, in both read and spontaneous TTS. Our work sheds light on both how speech SSL can readily improve current TTS systems, and how SSLs compare in the challenging generative task of TTS. Audio examples can be found at \href{https://www.speech.kth.se/tts-demos/ssr_tts}{https://www.speech.kth.se/tts-demos/ssr\_tts} 

% Hard character limit: 2000
\end{abstract} 
\section{Introduction}
\label{sec:intro}

Recently, text-to-speech (TTS) researchers have turned to using self-supervised learning (SSL) speech representations as intermediate representations in the popular two-stage TTS scheme \cite{siuzdak2022wavthruvec,du2022vqtts}, with an acoustic model followed by a vocoder.
These SSLs take the place of the conventionally used mel-spectrograms as acoustic-modelling targets.
% Similar to how word vectors \cite{mikolov2013efficient} and deep language models \cite{kenton2019bert} have reshaped text processing, 
They can be learned from audio alone (without transcriptions) thus can scale to vast amounts of data and compute to build better representations of speech audio.
They have demonstrated advantages across increasing spectrum of speech technology applications, with recent addition of building high-quality TTS from mixed-quality audio \cite{siuzdak2022wavthruvec}.
%in terms of .

However, SSL-based approaches to TTS are still in their infancy \cite{siuzdak2022wavthruvec, du2022vqtts, leehierspeech}.
Although speech SSLs have been compared in various applications such as ASR, speaker identification, and speech conversion \cite{yang21c_interspeech}, it remains unclear which SSL is superior for TTS and why.
%While speech SSLs have been compared in non-TTS applications (e.g., ASR, speaker identification, speech conversion \cite{yang21c_interspeech}), it is not clear which is the better choice for TTS and why.
% nor what impact choosing one or the other will have have.
The demonstrated effectiveness of intermediate layers of the same SSL model for downstream tasks \cite{yang21c_interspeech} further clouds the picture regarding what representation to use for TTS.
Moreover, two-stage TTS with SSL-based intermediate representations has thus far only been applied to speech audio of text read aloud \cite{siuzdak2022wavthruvec}, whereas most speech in the wild is spontaneous and unscripted.
Spontaneous speech has many unique verbal and nonverbal phenomena (e.g., breathing, disfluencies, discourse markers) often not represented in text, and is severely underrepresented in TTS research today due to these challenges.
SSLs have the potential to revolutionise spontaneous TTS as they do not require high-quality audio or transcripts to learn effectively, but to what extent they can live up to this potential is unknown.

We address these questions by building several two-stage TTS systems with different speech SSLs as intermediate representations, and study their objective resynthesis accuracies as well as subjective TTS qualities.\footnote{Audio examples: \href{https://www.speech.kth.se/tts-demos/ssr_tts}{https://www.speech.kth.se/tts-demos/ssr\_tts}.}
The key findings are:
\begin{enumerate}[(1)]
%\begin{enumerate}
\item Not all SSLs are created equal for TTS.
\item The final layer of a SSL fine-tuned for ASR is impoverished for TTS, compared to a middle layer.
\item The resynthesis error is not a good predictor of the overall TTS quality afforded by using a given SSL.
\end{enumerate}
We also present the first results on applying SSLs as intermediate representations in spontaneous TTS, where we find:
\begin{enumerate}[(1),resume]
\item Similar ordering of resynthesis and TTS performance within SSLs as in read speech.
\item SSL features outperform mel-spectrograms more in spontaneous TTS than in read-speech TTS.
% \item Reduced audio quality (due to the smaller amount of training data and/or the greater acoustic diversity of the material).
% \item As an informal observation, we find that the system inserts spontaneous-speech phenomena (e.g. disfluencies and discourse markers) not present in the input text.
\end{enumerate}
% A similar effect as the last point has been studied in \cite{szekely2019how}, but it has not been seen before with parallel TTS with monotonic alignments.

%In this work, we build two-stage spontaneous TTS systems with SSRs as intermediate representation, a first to the best of our knowledge.

\section{Background and prior work}
\subsection{Self-supervised speech representations}
Learning general-purpose representations from raw audio is a key question in speech processing. Borrowing from the success of self-supervised learning (SSL) in other domains (vision, language), speech SSL learned with only self-supervised objectives and then finetuned on down-stream tasks have shown promise in applications such as ASR and speaker identification \cite{yang21c_interspeech}. Examples of these representations are wav2vec \cite{schneider2019wav2vec}, wav2vec2.0 \cite{baevski2020wav2vec}, HuBERT \cite{hsu2021hubert}. Their initial success in ASR has led to research of their usage in generative tasks such as speech encoding/resynthesis \cite{polyak2021speech}, speech language modeling \cite{lakhotia2021generative}, and TTS \cite{siuzdak2022wavthruvec}.
%This work contributes to further exploring utilizing these SSRs for TTS.
% This work further explores the utility of these SSRs for TTS.

With the proliferation of SSL approaches, recent work has also focused on understanding the difference between SSL models \cite{yang21c_interspeech,ma2021probing} and the difference between layers of the same SSL model \cite{pasad2021layer,li2022exploration}. Our work extends these findings by testing different SSLs and their layers in two-stage TTS.

\subsection{Two-stage Text-to-Speech}
Neural speech synthesis today is dominated by a so-called ``two-stage" approach, where a stage-1/acoustic model first generates intermediate representations which are then vocoded by stage-2/vocoder model. Since the first successful neural two-stage TTS Tacotron2 \cite{shen2018natural}, the focus of research has been on improving either stage-1 or stage-2 architecture and/or training loss, e.g. transformer-based stage-1 model \cite{lancucki2021fastpitch, renfastspeech}, normalizing flow-based stage-1 model \cite{kim2020glow}, GAN-based stage-2 model \cite{kong2020hifi}. Only recently did the intermediate representation start to garner interest with approaches that use SSL instead of the default choice of mel-spectrogram \cite{siuzdak2022wavthruvec, du2022vqtts}. 
% It is still unclear which SSL approach is best suited for this setup, a question that we aim to investigate here. %But it is not clear which SSL is more suitable in this setup, a question we hope to address.

\subsection{Synthesizing spontaneous speech}
Modelling spontaneous speech is a critical part of creating more human-like speech synthesis \cite{tan2021survey}. 
However, synthesizing spontaneous speech is fundamentally different to read-speech synthesis which is more explored in TTS research today. The differences, which are mostly hindering factors for developing TTS, include more dynamic prosody, disfluencies, various breathing sounds, vocal outbursts, and large variations in voice and recording quality. Many of these characteristics are poorly approximated in the transcript or are completely left out \cite{szekely2019how}, making training TTS difficult. These challenges and spontaneous TTS as a whole are receiving increasing attention from TTS community \cite{szekely2019spontaneous, chen2023vector}.

\section{Method}
\label{sec:method}
\subsection{Representations}
Four speech SSLs were selected for testing, as summarized in Table \ref{tab:SSRs}. %Three different SSR models were included.
They are different in the training-loss design as well as in the data used for training. We also selected two different layers in the same model wav2vec2.0, as prior work has shown that a middle layer of a ASR-finetuned SSL model contains more prosodic information \cite{pasad2021layer} which could benefit synthesis. We recognize that this set is limited in some aspects and future work could extend by adding more comparisons that probe specific aspects. These SSLs are compared against standard 80-channel log mel-spectrogram used in current two-stage TTS.

\begin{table*}
\centering
\begin{tabular}{l|c|c|c|c|c|c|c}
\toprule
\textbf{Representation Notation}  & \textbf{Pre-training Data} & \textbf{ASR} & \textbf{Architecture} & \textbf{Layer} & \textbf{Hz} & \textbf{Dim} &\textbf{Resynthesis Error$\downarrow$} \\ 
\textbf{(SSL Model)} & \textbf{(Loss)}          & \textbf{Finetuning} &  &             & \textbf{} & & \textbf{Read$\bf{|}$Spontaneous} \\
\midrule
w2v2-L12 & LibriSpeech960H & LibriSpeech960H & 7CL$\rightarrow$ & TL-12 & 50 & 768 & 0.7115 $|$ 0.6724 \\
(wav2vec2.0-base-asr \cite{baevski2020wav2vec}) & (contrastive+diversity) &                 &  12TL     &  &  & & \\
\midrule
w2v2-L9 $\star$ & LibriSpeech960H & LibriSpeech960H & 7CL$\rightarrow$ & TL-9 & 50 & 768 & 0.6089 $|$ 0.5582 \\
(wav2vec2.0-base-asr \cite{baevski2020wav2vec}) & (contrastive+diversity) &                 &  12TL    &  &  & & \\
\midrule
w2v1-c  & LibriSpeech960H & - & 5CL(z)$\rightarrow$ & c & 100 & 512 & 0.5253 $|$ 0.4437 \\
(wav2vec \cite{schneider2019wav2vec})  & (contrastive) &  & 9CL(c) & & & &\\
\midrule
HuBERT-L12 & LibriLight60,000H &  LibriSpeech960H & 7CL$\rightarrow$ & TL-12 & 50 & 1024 & 0.4406 $|$ 0.4056\\
(HuBERT-large-asr \cite{hsu2021hubert}) & (masked prediction)  &                   &   24TL    & &  & &  \\ 
\bottomrule
\end{tabular}
\caption{\label{tab:SSRs} Speech SSL tested. In Architecture and Layer columns, ``CL" denotes convolutional layer and ``TL" denotes transformer layer. wav2vec2.0 and HuBERT share convolutional and transformer layer architecture. Vocoding Error column corresponds to Figure 2. $\star$: w2v2-L9 performs best in TTS perceptual test.}
\vspace{-1\baselineskip}
\end{table*}
% {
%     "LJS":{
%         "w2v2-L12": 0.7115,
%         "w2v2-L9": 0.6089,
%         "w2v1-c": 0.5253,
%         "HuBERT": 0.4406,
%         "mel-spec": 0.2453
%     },
%     "TSGD":{
%         "w2v2-L12": 0.6724,
%         "w2v2-L9": 0.5582,
%         "w2v1-c": 0.4437,
%         "HuBERT": 0.4056, 
%         "mel-spec": 0.1983
%     }
% }

\subsection{Data}
For read speech TTS, we used the widely benchmarked LJSpeech corpus \cite{ljspeech17}. For spontaneous TTS, we used the audio recordings of the Trinity Speech-Gesture Dataset (TSGD) \cite{IVA:2018}, comprising 25 impromptu monologues (avg. 10.6m long), %each on average 10.6 minutes long, 
spoken by a male actor in %an impromptu, 
colloquial style. %The corpus was transcribed using ASR and subsequently corrected manually.
To prepare the dataset for TTS, we segmented the corpus into stretches of speech delineated by breath events, and combined these %segments 
in an overlapping fashion to form an utterance structure. %to form utterances no longer than 11 seconds
\cite{szekely2020breathing}.  

\subsection{System}
\begin{figure}[h!]

% \begin{minipage}[b]{1.0\linewidth}
%   \centering
%   \centerline{\includegraphics[width=8.5cm]{image1}}
% %  \vspace{2.0cm}
%   \centerline{(a) Result 1}\medskip
% \end{minipage}

\centering
\includegraphics[width=\linewidth]{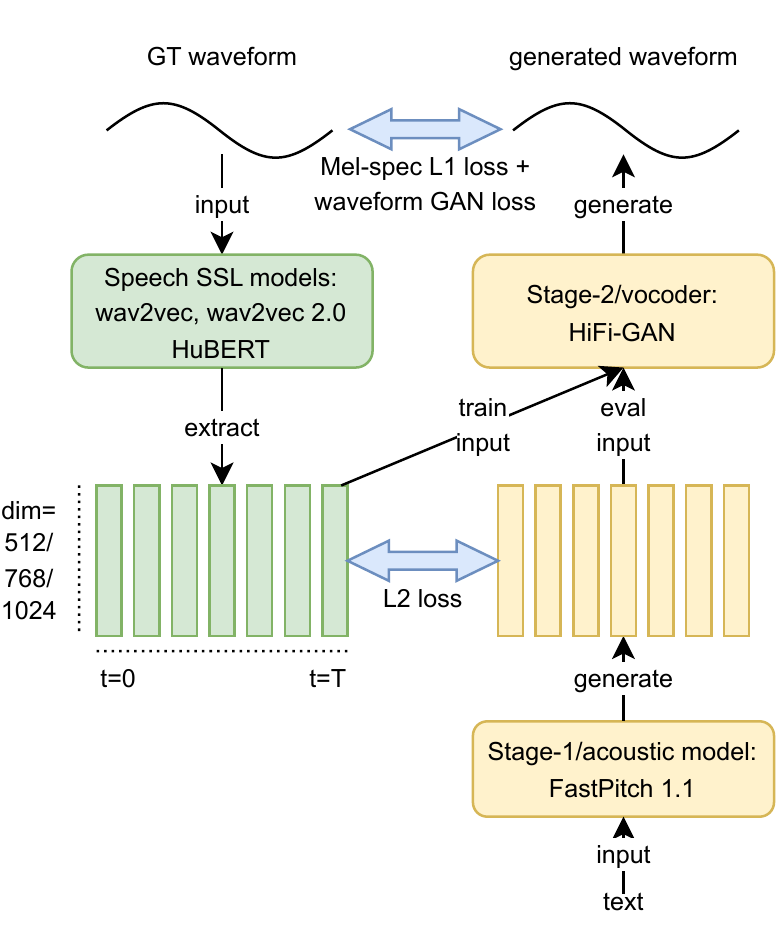}
\caption{Diagram of the system.}
% HiFi-GAN (stage-2/vocoder) is trained to reconstruct waveform from extracted SSLs, and a FastPitch model (stage-1/acoustic model) is trained to generate SSLs from text.}
\label{fig:main}
\end{figure}

We largely follow WavThruVec \cite{siuzdak2022wavthruvec} in neural architecture choice and training setup, both for simplicity and because architecture improvement is not the main focus of our work. Our architecture is essentially the single-speaker version of WavThruVec.  We also note that WavThruVec used wav2vec2.0-base-asr to demonstrate the effectiveness of the model, while we compare a range of SSL representations with the focus on probing their differences in WavThruVec framework.  An overview of the method is shown in Fig.\ \ref{fig:main}.

% TODO: introduce compared representations here

The stage-1 acoustic model is adapted from FastPitch \cite{lancucki2021fastpitch}, a parallel transformer-based TTS model, with added monotonic alignment \cite{badlani2022one}. \footnote{The model is called "Fastpitch 1.1" in the official implementation.} 
% It learns input-output alignment automatically through a monotonic alignment mechanism \cite{lancucki2021fastpitch}.
All stage-1 models are trained with identical hyperparameters as in \cite{siuzdak2022wavthruvec}, except with a batch size of 96. The models are trained for variable iterations until the validation loss stops improving depending on the corpus, which takes about 200 epochs for most models on LJSpeech. We use an official pre-trained FastPitch \cite{lancucki2021fastpitch} as mel-spec baseline.

During initial experiments with the spontaneous corpus TSGD, we found that random weight initialization results in low-quality speech. This is likely due to the smaller size of this corpus, at approximately 4 hours in total. To obtain better results, we instead start training from our LJSpeech-trained first-stage model and finetuned it on the spontaneous corpus for 200 epochs. This methodology is highly effective in allowing neural TTS to be trained on smaller databases \cite{szekely2019spontaneous}.

The stage-2 model (vocoder) is adapted from HiFi-GAN \cite{kong2020hifi}, trained in a similar setup as in \cite{siuzdak2022wavthruvec}. Training batches consist of 0.5s random audio excerpts, with the batch size varying from 160 to 192 depending on the dimension and sampling rate of the representation, to ensure that per-iteration data throughput is approximately consistent across models. All LJSpeech SSL vocoders are trained for 180k iterations at 22 kHz and we use an official pre-trained LJSpeech mel-spectrogram HiFi-GAN from authors of FastPitch \cite{lancucki2021fastpitch} as baseline. All vocoders for spontaneous corpus \cite{IVA:2018} are trained for 100k iterations at 16 kHz from scratch.

\section{Results}
\label{sec:results}

\subsection{Resynthesis}
\begin{figure}[b!]
\vspace{-2em}
\centering
\includegraphics[clip, trim=0cm 0.2cm 0cm 0.7cm, width=\linewidth]{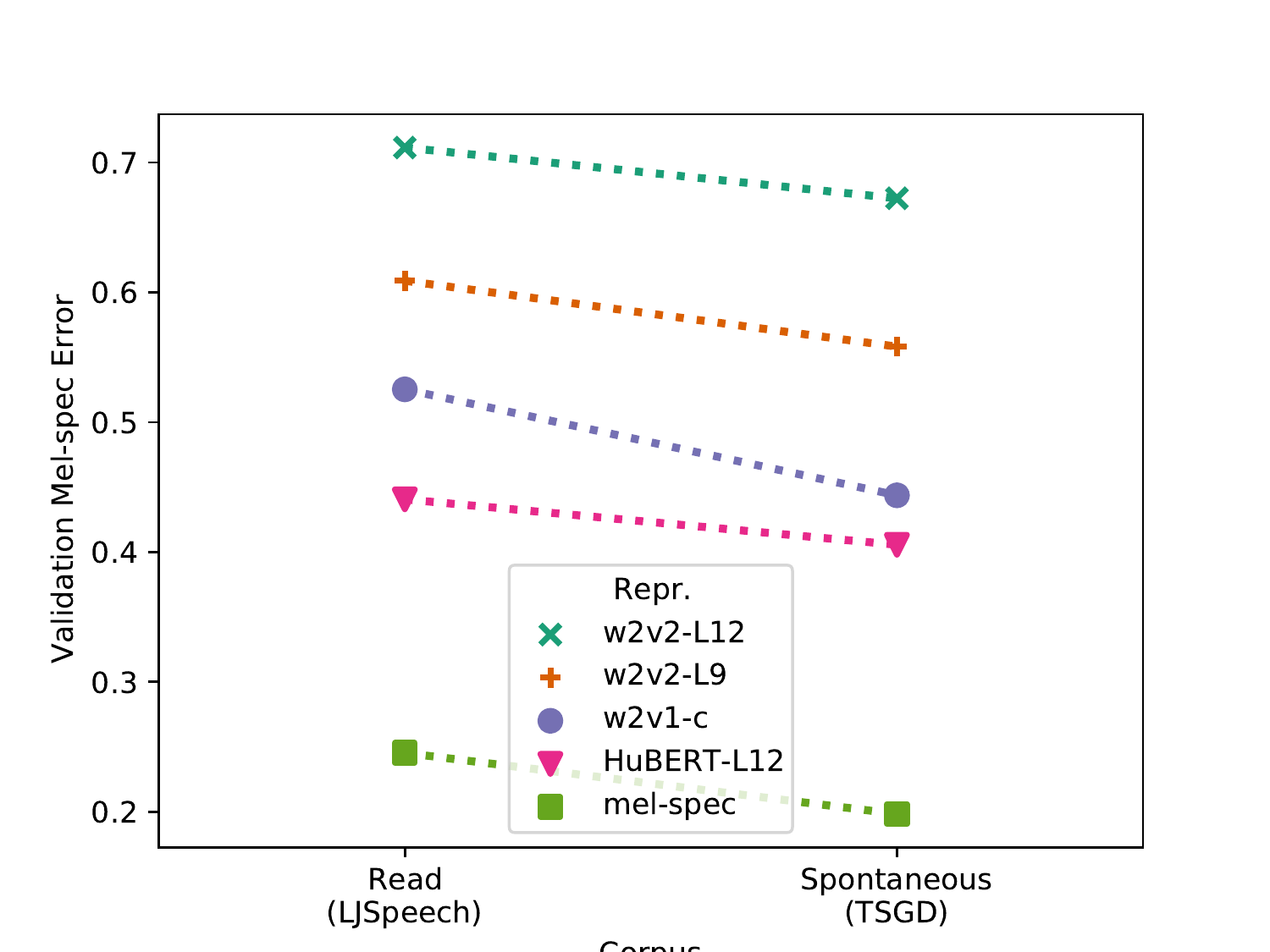}

\caption{Resynthesis performance comparison between representations on the two corpora.}
\label{fig:resynthesis}
\end{figure}

% We train HiFi-GAN on the two corpora, read-speech LJSpeech (LJS) \cite{ljspeech17} and spontaneous-speech \cite{IVA:2018}. The representations are extracted from original waveforms and HiFi-GAN is trained to reconstruct the waveforms from the extracted representations. 
We first look at resynthesis performance of the SSL representations. We extract SSLs from speech audios in the validations set and then reconstruct the waveforms using the SSLs' respectively trained Hifi-GAN vocoders. Mel-spectrogram error (L1) of the reconstructed waveform is used as a measure of resynthesis performance. The results are visualised in Fig.\ \ref{fig:resynthesis} with corresponding values listed in the last column of Table \ref{tab:SSRs}.

The ordering of representations is consistent in both read and spontaneous corpora resynthesis performance. Between the two wav2vec2.0 layers, L9 has better resynthesis performance than L12. This is consistent with prior findings that earlier layers of wav2vec2.0-ASR contain more prosodic information than later layers \cite{pasad2021layer} \cite{li2022exploration}. 
% However, we are the first work to show this from resynthesis point of view. 

\subsection{Text-to-speech}
We conducted comparison-MOS (CMOS) listening tests to compare TTS performance of the SSLs. Each test consists of two speech audio clips synthesized by two models with the same input text, and the listener is asked to choose between 3 scales of preferences towards either audio or a no-preference, totaling 7 options. The scores thus range from [-3, 3]. \footnote{This setup of CMOS is as described in ITU-T P808 (“Subjective evaluation of speech quality with a crowdsourcing approach”).} We conducted all listening tests on the crowdsourcing platform Prolific. Each set of CMOS test was completed by a separate group of at least 30 participants of native English speakers.

\begin{table}[t!]
\centering
\begin{tabular}{@{}ll|r|r@{}}
\toprule 
                  &                         & \multicolumn{2}{c}{CMOS (pref.\ for Rep.\ 1)}\tabularnewline
 \multicolumn{2}{c|}{}                      & \multicolumn{1}{c|}{Read} & \multicolumn{1}{c}{Spontaneous} \tabularnewline
  \multicolumn{2}{c|}{Rep.\ 1 vs.\ Rep.\ 2} & \multicolumn{1}{c|}{(LJSpeech)} & \multicolumn{1}{c}{(TSGD)} \tabularnewline
\midrule
 w2v2.0-L9 & w2v2.0-L12 & 1.007* & 0.626*\tabularnewline
 w2v2.0-L9 & HuBERT-L12     & 0.793* & 1.012*\tabularnewline
 w2v2.0-L9 & mel-spec.  & 0.354* & 1.877*\tabularnewline
 HuBERT-L12 & mel-spec.     & -0.614* & - \tabularnewline
\bottomrule
\end{tabular}
\caption{Comparison-MOS (CMOS) Results from TTS subjective listening tests (*: p-value obtained from Wilcoxon signed-rank test $<$ 0.001).}
\label{tab:cmos}
\end{table}

\subsubsection{Read-speech TTS}
\label{sec:ljs_results}
We found through subjective listening that w2v2-L9 is superior in read-speech TTS than other representations, thus we conducted four sets of CMOS tests to confirm this finding as shown in the LJSpeech column of Table \ref{tab:cmos}.  We did not test w2v1-c in listening tests due to its low output intelligibility and overall bad synthesis quality that is clearly much inferior than other representations. The test input are 40 meaning-neutral sentences chosen from the test set of the corpus prior to testing. 
% \begin{enumerate}[(1)]
% \item w2v2-L9 and w2v2-L12, testing that a middle layer of w2v2 indeeed is a better fit for TTS on a read TTS corpus,
% \item w2v2-L9 and mel-spectrogram,
% \item HuBERT and mel-spectrogram.
% \end{enumerate}
% Tests (2), (3) benchmark using a SSR against mel-spectrogram in a two-stage TTS pipeline, where other conditions are similar (TTS model, vocoder, training setup). We did not test w2v1-c in listening tests due to its clearly inferior synthesis quality.
The CMOS results (Table \ref{tab:cmos}) show that w2v2-L9 outperforms all other SSLs and mel-spec. HuBERT-L12 does not perform better than mel-spec. 
These results show that, in a standard read TTS corpus, using SSL can perform better than using mel-spectrogram as representation medium in a two-stage TTS pipeline. However, the performance of different SSLs also differ greatly.

\subsubsection{Spontaneous TTS}
We conducted same set of CMOS tests as in read-speech TTS to confirm that w2v2-L9 is also superior in spontaneous TTS. We did not test w2v1-c in this setting due to its clearly low output quality similar to that in read-speech TTS.

The original test set from the spontaneous corpus contains largely context-dependent utterances which makes sentence-wise evaluation difficult. We instead fine-tuned a pre-trained generative language model on the training set and used it to generate 20 independently coherent sentences.

The results are shown in TSGD column in Table \ref{tab:cmos}. w2v2-L9 again outperforms all other SSLs and mel-spec. This is consistent with results from read speech TTS (Sec.\ \ref{sec:ljs_results}).
% This shows that, in a relatively small spontaneous corpus, SSL can perform much better than mel-spectrogram as representation medium in a two-stage TTS pipeline. 
Notably the preference for w2v2-L9 over mel-spec is much higher in spontaneous TTS than in read TTS, suggesting that SSL could be especially suitable as intemediate representation in two-stage spontaneous TTS. 

% We also observed that TTS with SSL (w2v2-L9 and HuBERT-L12) as intermediate representations sporadically inserts fillers (`um', `uh') by itself. This phenomenon has only been shown in a stochastic stage-1 model like Tacotron 2 \cite{szekely2019how}, while the stage-1 model in our system, FastPitch, is deterministic. It is thus likely due to the SSL used, and subject to further investigation in future work.

\section{Discussion}
It is clear from Table \ref{tab:SSRs} and \ref{tab:cmos} that resynthesis performance does not predict overall TTS performance. This suggests that the better SSL for two-stage TTS is not necessarily the one that contains more acoustic information. If we look at other aspects of the tested SSLs as summarized in Table \ref{tab:SSRs}, it is also hard to find one single axis of variation that fully explains either resynthesis or TTS performance ranking.  A wider comparison with other SSL models and more layers of the same SSL could uncover which SSL and which layer is most suitable for TTS. Another potential addition to further comparison is the use of weighted \cite{yang21c_interspeech} or stacked SSL layers.

From the perspective of speech SSL research, our study shows that TTS (in both read and spontaneous speech settings) is another feasible downstream task, the performance of which could be taken into consideration when evaluating speech SSLs. However, we note that, (1) the specific TTS models could have bias towards which SSLs they work better with, (2) unlike other more easily benchmark-able downstream tasks, TTS is known to be complex to evaluate quantitatively or subjectively \cite{wagner2019speech}. But regarding the second point, we also note that SSL itself has already emerged as a promising tool for automatic evaluation of TTS systems \cite{cooper2022generalization}.

\section{Conclusions}
\label{sec:conclusions}

We compared several speech SSLs as the representation medium in a standard two-stage TTS pipeline, on both read and spontaneous speech corpora. We used the same models and training setup for both stages of the TTS pipeline for fair comparison. Listening tests were conducted in comparative-MOS (CMOS) form. The results show that one tested SSL, layer 9 of 12-layer wav2vec2.0-ASR, outperforms other tested SSLs as well as the mel-spectrogram on both read and spontaneous TTS. Our results also show how utilizing SSL in TTS is a promising new research direction, and that it effectively addresses challenges in spontaneous speech synthesis.
%in spontaneous speech synthesis both due to both the data quality-related and the expressive variability-related challenges of spontaneous speech synthesis.
%Future work will be devoted to further optimizing SSR towards the spontaneous TTS task.
Interesting directions for future work include wider comparisons of SSLs in TTS and further optimizing SSL towards the spontaneous TTS task.

\bibliographystyle{IEEEbib}
% \bibliography{strings,refs,main}
\bibliography{main}

\end{document}